%% file: main.tex
\documentclass[11pt]{article}

\def\ShowComment{True} 

\input{header}

\input{macro}
\title{Income Inequality, Food Aid, and 'Zero Hunger': Evaluating Effectiveness During Lula's Administration}
\author[1]{Bo Wu}
\affil[1]{Beijing International Studies University}

\date{}

\begin{document}

\maketitle
\pagenumbering{gobble}
\input{sections/abstract}

\newpage
\pagenumbering{arabic}

\input{sections/intro}
\input{sections/prelim}
\input{sections/static}

\input{sections/dynamic}

\appendix

\newpage
\bibliographystyle{alpha}
\bibliography{main.bib}

\end{document}

%% file: header.tex
\usepackage[utf8]{inputenc}
\usepackage[T1]{fontenc}

\usepackage{authblk}

\usepackage{subfigure} 
\usepackage{graphicx}  

\usepackage{geometry}
\usepackage[bottom]{footmisc}
\usepackage{todonotes}
\usepackage{xspace}

\usepackage{comment}

\usepackage{nicefrac}
\usepackage[noend]{algpseudocode}

\usepackage{amssymb}
\usepackage{amsmath}
\usepackage{amsthm}
\usepackage{dsfont}
\usepackage{footnote}

\usepackage{array}
\usepackage{verbatim}
\usepackage{multirow}
\usepackage{xcolor}
\usepackage{nameref}
\definecolor{ForestGreen}{rgb}{0.1333,0.5451,0.1333}
\definecolor{DarkRed}{rgb}{0.65,0,0}
\definecolor{Red}{rgb}{1,0,0}
\usepackage[linktocpage=true,
pagebackref=true,colorlinks,
linkcolor=DarkRed,citecolor=ForestGreen,
bookmarks,bookmarksopen,bookmarksnumbered]
{hyperref}
\usepackage{cleveref}

\usepackage{thm-restate}
\usepackage[tableposition=bottom]{caption}
\usepackage[all]{hypcap}
\usepackage{subcaption}
\usepackage{framed}
\usepackage[framemethod=tikz]{mdframed}
\usepackage[skins]{tcolorbox}

\usepackage[ruled,vlined,linesnumbered]{algorithm2e}
\Crefname{algocf}{Algorithm}{Algorithms}

\makeatletter
\g@addto@macro{\maketitle}{\@thanks}
\makeatother

\newcommand{\p}{\textsc{P}}%

\renewcommand{\algorithmiccomment}[1]{\bgroup\hfill$\rhd$~#1\egroup}

\newcounter{note}[section]

\renewcommand{\paragraph}[1]{\medskip\noindent\textbf{#1}}

\ifdefined\ShowComment
\def\thatchaphol#1{\marginpar{$\leftarrow$\fbox{T}}\footnote{$\Rightarrow$~{\sf\textcolor{ForestGreen}{#1 --Thatchaphol}}}}

\def\sayan#1{\marginpar{$\leftarrow$\fbox{S}}\footnote{$\Rightarrow$~{\sf\textcolor{purple}{#1 --Sayan}}}}

\def\peter#1{\marginpar{$\leftarrow$\fbox{P}}\footnote{$\Rightarrow$~{\sf\textcolor{green}{#1 --Peter}}}}
\else
\def\thatchaphol#1{}
\def\sayan#1{}
\def\peter#1{}
\fi 

\usepackage{etoolbox}
\usepackage{tikz}
\usetikzlibrary{tikzmark}
\usetikzlibrary{calc}

\errorcontextlines\maxdimen

\newcommand{\ALGtikzmarkcolor}{black}
\newcommand{\ALGtikzmarkextraindent}{4pt}
\newcommand{\ALGtikzmarkverticaloffsetstart}{-.5ex}
\newcommand{\ALGtikzmarkverticaloffsetend}{-.5ex}
\makeatletter
\newcounter{ALG@tikzmark@tempcnta}

\newcommand\ALG@tikzmark@start{%
	\global\let\ALG@tikzmark@last\ALG@tikzmark@starttext%
	\expandafter\edef\csname ALG@tikzmark@\theALG@nested\endcsname{\theALG@tikzmark@tempcnta}%
	\tikzmark{ALG@tikzmark@start@\csname ALG@tikzmark@\theALG@nested\endcsname}%
	\addtocounter{ALG@tikzmark@tempcnta}{1}%
}

\def\ALG@tikzmark@starttext{start}
\newcommand\ALG@tikzmark@end{%
	\ifx\ALG@tikzmark@last\ALG@tikzmark@starttext
	\else
	\tikzmark{ALG@tikzmark@end@\csname ALG@tikzmark@\theALG@nested\endcsname}%
	\tikz[overlay,remember picture] \draw[\ALGtikzmarkcolor] let \p{S}=($(pic cs:ALG@tikzmark@start@\csname ALG@tikzmark@\theALG@nested\endcsname)+(\ALGtikzmarkextraindent,\ALGtikzmarkverticaloffsetstart)$), \p{E}=($(pic cs:ALG@tikzmark@end@\csname ALG@tikzmark@\theALG@nested\endcsname)+(\ALGtikzmarkextraindent,\ALGtikzmarkverticaloffsetend)$) in (\x{S},\y{S})--(\x{S},\y{E});%
	\fi
	\gdef\ALG@tikzmark@last{end}%
}

\apptocmd{\ALG@beginblock}{\ALG@tikzmark@start}{}{\errmessage{failed to patch}}
\pretocmd{\ALG@endblock}{\ALG@tikzmark@end}{}{\errmessage{failed to patch}}
\makeatother

%% file: sections/abstract.tex
\begin{abstract}
Income inequality has been an important social issue that has attracted widespread attention. Taking the Zero Hunger Program in Brazil as a case study, this study analyzes the impact of policy changes on the income distribution of the Brazilian population during the implementation of the program using a breakpoint regression approach. The data for the study come from a variety of sources, including the Brazilian Ministry of Development, Social Assistance and the Family, and are analyzed using detailed descriptive statistics from the CEIC Brazil In-Depth Database. The results of the study show that during the Lula administration, Brazil's Zero Hunger Program substantially reduced income inequality, provided more substantial income security for the poor, and reduced the income gap between the rich and the poor. In terms of gender differences, the program led to a larger increase in the income of the male labor force at the right age compared to the female labor force, further highlighting the positive impact of the policy on the male labor force. These results are further confirmed by sensitivity analysis and provide useful lessons for subsequent policy formulation. This study's deep dive into the effects of the Zero Hunger program provides a valuable contribution to academic social science research and policy development.


\end{abstract}

%% file: sections/intro.tex
\section{Introduction}
In the second half of the 20th century, Brazil experienced a series of political and economic upheavals, including military dictatorship and high inflation, which further exacerbated income inequality. During this period, the gap between the rich and the poor widened dramatically, plunging large segments of the population into poverty and extreme poverty.

Luiz Inácio Lula da Silva, commonly known as Lula, served as Brazil's president from 2003 to 2010 and introduced the "Zero Hunger Program" (Bolsa Família). This initiative aimed to assist poor families through cash transfer payments and other social support measures, promoting more equitable and inclusive economic growth\cite{cattaneo2016inference}. The program achieved significant success during Lula’s tenure, markedly reducing income inequality. Following its implementation, the quality of life for the impoverished improved, and the income gap between the rich and the poor narrowed.

However, over time, Brazil faced new challenges, including global economic instability, political corruption, and rising social discontent. These issues severely impacted Brazil’s economy and society, placing renewed pressure on the problem of income inequality.

Thus, Brazil’s economic and social context is highly complex, characterized by longstanding wealth disparities and income inequality, as well as the partial success of the "Zero Hunger Program" implemented during Lula’s presidency. These factors collectively shape Brazil’s current socioeconomic landscape\cite{silva2011zero}. To address these issues, the Brazilian government launched the "Zero Hunger Program" in 2003, aiming to eradicate hunger and poverty within the country. By providing cash transfers and food assistance, the program alleviated the economic burden on the poor and promoted social equity. However, its actual impact remains a subject of debate, particularly regarding its effects on income inequality and food security.

The Brazilian government’s social assistance program, "Zero Hunger," has garnered significant attention. Marques et al. highlighted the program’s success in eliminating hunger and its progress toward sustainable development (Marques, GaneM, Júnior, 2013.12). Meanwhile, Meade et al., through an assessment of food security and access, uncovered the intrinsic link between hunger and insufficient income (Meade, Valdes, Rosen, 2013, 5). Miller’s research focused on the Canela indigenous community, exploring the impact of social assistance programs, particularly Bolsa Família, on women and children (Miller, 2013, 6). Graziano da Silva et al. emphasized the maturity of the "Zero Hunger" proposal and its necessity as a national priority (Graziano da Silva, 2011, 11). Suplicy, analyzing the commitments and actions of Lula’s administration, examined Brazil’s trend toward achieving a basic citizen’s income (Suplicy, [6]).

To evaluate the policy’s actual effects, this study employs a regression discontinuity approach to deeply explore its potential impact on income distribution among Brazilian residents\cite{paes2014zero}. The study utilizes publicly available data from Brazil’s Ministry of Development, Social Assistance, and Family, as well as the CEIC Brazil Premium Database, combined with government survey reports and datasets from relevant research institutions, to conduct detailed descriptive statistics and analysis of income distribution.

Limited by data availability, this study references a set of rdlocrand Stata command packages developed by Professor Cattano of Princeton University and collaborators in 2016, employing randomized experimental methods to address the issue of insufficient effective sample size (i.e., the number of samples within the bandwidth). Using regression discontinuity, the study analyzes and evaluates the effectiveness of the "Zero Hunger Program" while also reflecting on President Lula’s inaugural speech—specifically, whether the program’s impact varied due to changes in government administration\cite{meade2004brazil}. The findings indicate that during Lula’s administration, the "Zero Hunger Program" substantially reduced income inequality, provided better income security, and narrowed the wealth gap. However, after Lula’s term ended, while income inequality continued to decline, the rate of reduction slowed, and the income gap stabilized, ceasing to shrink further.

%% file: sections/prelim.tex
\section{Preliminaries}
\subsection{Application of Regression Discontinuity Design (RDD)}
To thoroughly evaluate the impact of the "Zero Hunger Program" on income inequality in Brazil, this study adopts the Regression Discontinuity Design (RDD) method. The "Zero Hunger Program," a flagship social policy during Lula’s administration, aimed to alleviate poverty and food insecurity. RDD, as a robust econometric approach, is particularly suited for analyzing natural experiments or policy-induced discontinuities, enabling a precise examination of the potential causal effects of policy changes during the implementation phase of the "Zero Hunger Program" on residents’ income distribution. In this study, we focus on two key aspects of income inequality: gender disparities and the wealth gap\cite{kilpatrick2011fighting}. By observing changes in labor income between men and women, as well as shifts in the rich-poor divide before and after the program’s implementation, we infer the policy’s impact on Brazil’s income structure. This design allows us to more reliably assess the potential causal effects of the "Zero Hunger Program," avoiding misleading interpretations based solely on simple correlations.
\subsection{RD Design Under the Local Randomization Assumption}
Since classical RDD requires a large sample size, both parametric and nonparametric estimations demand that the density of the assignment variable be continuous at the cutoff point. Moreover, nonparametric estimation, which relies on local polynomial methods for statistical inference, is grounded in large-sample asymptotic theory. This poses a challenge: if the bandwidth is too narrow or the dataset itself is small, the smoothness test may fail, or the estimation quality may deteriorate\cite{miller2013hunger}. To address the issue of insufficient effective sample size (i.e., the number of samples within the bandwidth), this study employs the rdlocrand Stata command package, developed in 2016 by Professor Cattano of Princeton University and his collaborators, which uses randomization-based methods.

This approach is based on a regression discontinuity design under the local randomization assumption. The rdlocrand package, designed by Professor Cattano and his team, includes four commands tailored for finite-sample inference in RD designs. The first command, rdrandinf, uses randomization methods to infer point estimates, hypothesis tests, and confidence intervals based on various assumptions\cite{mcclennen2011aesthetics}. The second command, rdwinselect, selects a window near the cutoff where the assumption of random treatment assignment is most plausible. The third command, rdsensitivity, conducts a series of hypothesis tests across different windows around the cutoff to assess the method’s sensitivity and construct confidence intervals. Finally, the fourth command, rdrbounds, provides sensitivity analysis for the implemented RD design based on randomization. The rdlocrand package offers a comprehensive framework for analyzing RD designs under the local randomization assumption and provides alternative finite-sample methods suitable for empirical analysis.

Thus, this study leverages this methodology to conduct the empirical analysis presented herein.

%% file: sections/static.tex
\section{The Static Algorithm}
\input{sections/tables}
\subsection{Data Sources and Variable Definitions}
The data for this study is sourced from publicly available datasets provided by Brazil’s Ministry of Development, Social Assistance, Family, and Hunger Eradication, as well as the CEIC Brazil Premium Database. We collected socioeconomic data and income distribution data from before and after the implementation of the program, supplemented by relevant reports, including but not limited to survey reports issued by government agencies and datasets from related research institutions, to ensure the reliability and completeness of the data.

Additionally, it is necessary to clarify the program and the associated data. For the "Zero Hunger Program," the Lula administration primarily implemented two measures: the Food Card Program (enabling families to purchase food) and the Food Acquisition Program (a transfer payment model known as PAA through government procurement). However, in 2004, to advance the "Zero Hunger Program," most of these components were integrated into the Family Assistance Program (Plan of Bolsa Família). Consequently, when selecting data for this study, we opted for data starting from 2004 and chose the number of families receiving assistance as the covariate for the regression discontinuity analysis. To fully reflect regional variations, we gathered data from all 26 Brazilian states plus the Federal District, totaling 27 groups for analysis.

Regarding gender differences, we selected the average annual monthly income of male workers and female workers within the labor force across the 27 regions as the dependent variables. These variables are used to assess the impact of the "Zero Hunger Program" on labor income across Brazilian regions during Lula’s administration. For income inequality, we adopted the Gini coefficient for each region as the measurement indicator. The Gini coefficient is a widely used metric for assessing income distribution inequality, ranging from 0 to 1, where a higher value indicates greater income inequality and a lower value signifies more equitable income distribution. Through these two indicators, we examine the program’s impact on promoting citizens’ income growth from both a horizontal perspective (gender) and a vertical perspective (wealth gap). The study spans from 2004 to 2015, with the cutoff point set at Lula’s departure from office in 2011. For statistical convenience, we define 2011 as the baseline (margin = 0), meaning 2004 is coded as -7, 2003 as -6, and so forth.

The selection of variables is in this \textbf{table~\ref{variables}}.
\subsection{Empirical Preparation}
First, based on the methodology of this study, we conducted window selection with the following key steps:

\paragraph{Window Setting}
According to the empirical design of this study, we aim to evaluate the impact of the "Zero Hunger Program" on income inequality during Lula’s administration. Thus, the cutoff point is set at the end of Lula’s term, specifically January 1, 2011. Before and after this cutoff, there are 6 time points (pre-cutoff) and 4 time points (post-cutoff), respectively.

From the results of descriptive statistics, we calculated the maximum value of the margin as 5 and the minimum value as -7. Therefore, the largest window is set as [-5, 5]. Within the window [-1, 1], there are 6 values on the left side of the cutoff and 5 values on the right side, making the smallest window [-1, 1].

\input{figures/pvalue}
\paragraph{Setting Window Increments and Covariates}
In this study, we used an increment of 0.125. Across all symmetric windows near the cutoff between [-1, 1] and [-5, 5], we applied the randomization-based exact RDD null hypothesis test to the covariates.

Based on the above settings, covariate testing begins with the smallest window [-1, 1] and increases incrementally by 0.125. The rdwinselect command generates \textbf{Figure~\ref{pvalue}} and recommends the window [-3.0, 3.0].

\input{figures/result}
\subsection{Empirical Process}
Based on the recommended window provided earlier, we conducted a regression discontinuity analysis, yielding the \textbf{Figure~\ref{pvalue}}:

Among the variables, the monthly income of working-age women and the wealth gap were statistically significant at the 5\% level, while the monthly income of working-age men was significant at the 10\% level.

\textbf{Figure~\ref{result}} presents the regression discontinuity plots of the explanatory variable (PBF) against the three dependent variables selected for this study. From the figure, we can observe that the income of Brazil’s working-age labor force has experienced a steady increase. Notably, the income of male workers, both in terms of growth rate and magnitude, surpasses that of female workers. This suggests that the policy disproportionately benefited male workers. Additionally, prior to the cutoff, the income of both male and female workers exhibited an upward trend. However, after the cutoff, while wage growth continued, the pace slowed over time.

This pattern is similarly reflected in the wealth gap. After the end of Lula’s administration, the decline in the Gini coefficient began to slow, indicating that the rate of reduction in income disparity among Brazilians decreased and stabilized.

The regression discontinuity analysis reveals that the "Zero Hunger Program," initiated during Lula’s tenure, had a substantial impact on income inequality. First, the program significantly reduced income inequality, providing better income security for the poor. By comparing the income gap before and after the program’s implementation, we observe that the disparity between the rich and the poor narrowed following its introduction. This suggests that the benefits of the "Zero Hunger Program" genuinely reached disadvantaged groups, offering them greater opportunities to access equitable economic resources.

%% file: sections/tables.tex

\begin{table}[htbp]
\centering
\small  
\resizebox{\textwidth}{!}{  
\begin{tabular}{llll}
\hline
\textbf{Variable Type} & \textbf{Variable Name} & \textbf{Definition} & \textbf{Unit} \\ \hline
Dependent Variable & Gini & Indicator measuring the degree of income inequality & / \\ 
 & Valmaleincoming & Monthly wage of working-age men & BRL (Real) \\ 
 & Valfemaleincoming & Monthly wage of working-age women & BRL (Real) \\ \hline
Independent Variable & PBF & Households receiving assistance from the Zero Hunger program in \textbf{state i in year t} (Plan of Bolsa Família, PBF) & Household(s) \\ \hline
Driving Variable & Margin & Difference from the base year & / \\ \hline
\end{tabular}
}
\caption{\textbf{Variable Definitions and Units}}
\label{variables}
\end{table}

%% file: figures/pvalue.tex
\begin{figure}[htpb]
	\centering
	\hspace{-1.5mm}
	\includegraphics[width=0.8\linewidth]{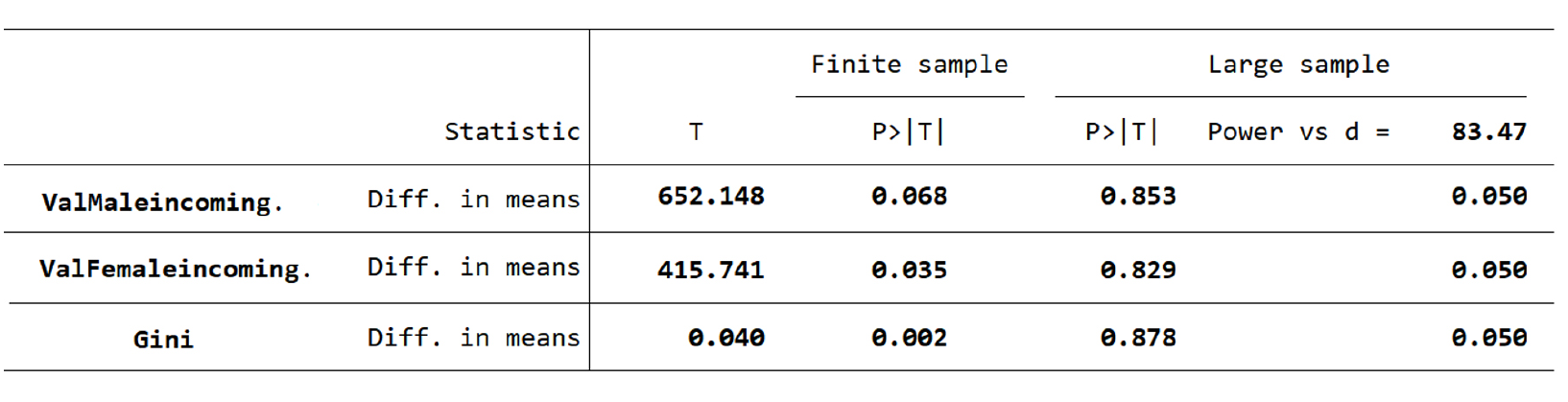}
	\caption{\textbf{The p-value of the regression discontinuity of the three explained variables}}
	\label{pvalue}
	\vspace{-1.5mm}
\end{figure}


%% file: figures/result.tex
\begin{figure}[htpb]
	\centering
	\hspace{-1.5mm}
	\includegraphics[width=0.8\linewidth]{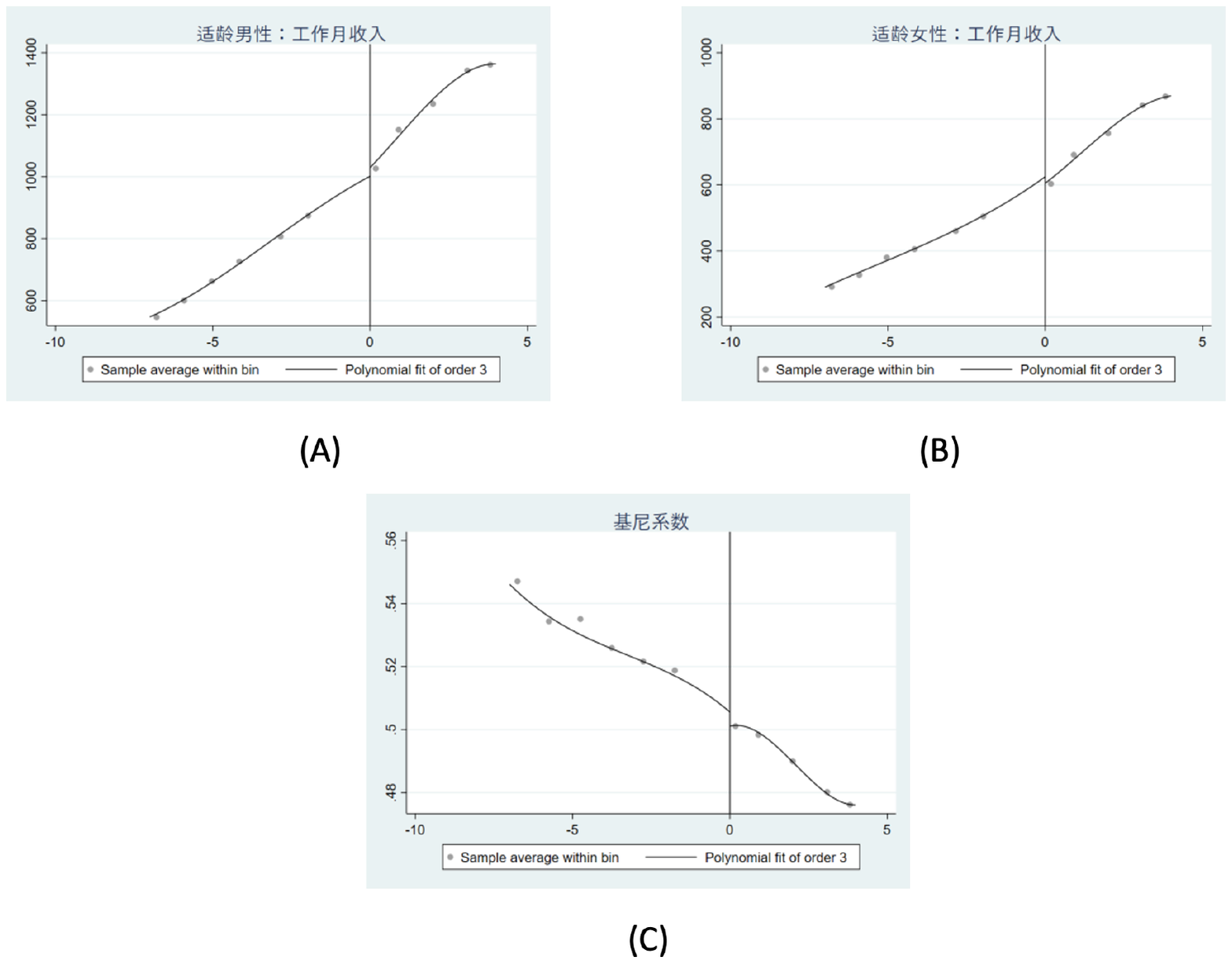}
	\caption{\textbf{Regression results of the explanatory variable PBF on the three explained variables}}
	\label{result}
	\vspace{-.5mm}
\end{figure}


%% file: sections/dynamic.tex
\section{Sensitivity Analysis}
\input{figures/winx}
\subsection{Robustness Analysis of Window Size}
For the three dependent variables, we examine the sensitivity of the results from two perspectives.

First, we analyze the impact of window size on robustness. As shown in the right panel of the figure, the results from the first row of commands reflect changes across four windows. With varying window sizes, we observe that in the column corresponding to 2.875, the values range from 90 to 200 without rejecting the null hypothesis. This suggests that within this window, there is a certain treatment effect. However, as the window continues to change, all points remain highly robust. Thus, at the same confidence level, the results under treatment remain relatively stable even as the bandwidth varies.

The results for the other two windows follow a similar pattern. For female workers, at a 95\% significance level, the treatment effect in the window [-2.875, 2.875] ranges from 90 to 200. However, in the windows [-3, 3], [-3.125, 3.125], and [-3.25, 3.25], the results remain consistently robust. Therefore, we conclude that the sensitivity of the treatment effect in this segment is relatively low, and the results are robust. For the Gini coefficient, the results are entirely robust.

These findings are illustrated in \textbf{Figure~\ref{winx}}. From the perspective of confidence intervals, the conclusions appear highly reliable. However, relying solely on the values of the treatment effect confidence intervals does not provide a fully dependable conclusion. Further analysis based on different test statistics is presented below.

\input{figures/robust}
\subsection{Robustness Test with Different Test Statistics}
\textbf{Figure~\ref{robust}} presents the results for the three dependent variables under different test statistics as the window size changes. When the window width is fixed, random experiments with varying degrees of deviation do not affect the p-values across the three dependent variables. This indicates that, as the window size changes, the results of the test statistics show no significant differences. Therefore, for the model-recommended bandwidth [-3.0, 3.0], the results remain robust across various levels of skewness.

\section{Summary and Limitations}
This article discusses the impact of Brazil's "Zero Hunger" program, launched by the Lula government in 2003, on income inequality. It employs the regression discontinuity method to analyze how policy changes during the Lula administration affected income distribution. The study utilizes multiple data sources, including descriptive statistics and analysis from the Brazilian Ministry of Development, Social Assistance and Family, and the CEIC Brazil Premium Database\cite{suplicy2003president}. The findings indicate that the program significantly reduced income inequality during the Lula government, providing better income security for the poor, narrowing the income gap between the rich and the poor, and that government transitions also had some influence on the policy’s effectiveness. Additionally, the study notes that the program had a more positive impact on the income growth of working-age male workers compared to their female counterparts. These results were further validated through sensitivity analysis.

However, we must acknowledge that this study is not without certain limitations. First, the method relies on the assumption that a structural change occurs at a specific point, and the selection of this point involves the researcher’s subjective judgment, which may introduce personal bias.

Second, regression discontinuity often assumes that structural changes are uniform across the entire sample, but in reality, heterogeneity may exist across different subgroups. Moreover, since this study adopts the research design method of Cattaneo from Princeton University, it assumes local randomization for small sample sizes—namely, that near the discontinuity point, the treatment variable is assigned with a high probability of randomization. However, for smaller samples or irregularities in distribution, this assumption may not hold, potentially affecting the validity of the estimation results.

Furthermore, to satisfy the local randomization assumption, an appropriate window size must be chosen to ensure the treatment variable near the discontinuity point exhibits random properties. However, the choice of window size can be subjective, and an improper selection may lead to biased estimation results.

Finally, this study is constrained by factors such as the incompleteness of available data and focuses solely on the family assistance program as the only covariate. Other factors, such as training, education, and the positive effects of government public works, may represent unconsidered variables that could influence the results. In future research, alternative methods or the inclusion of additional control variables could be employed to address these issues, yielding more accurate and reliable conclusions.

%% file: figures/winx.tex

\begin{figure}[htpb]
	\centering
	\subfigure[L-Monthly wages of working-age male workers change with window R-Monthly wages of working-age female workers change with window]{
		\includegraphics[width=0.8\linewidth]{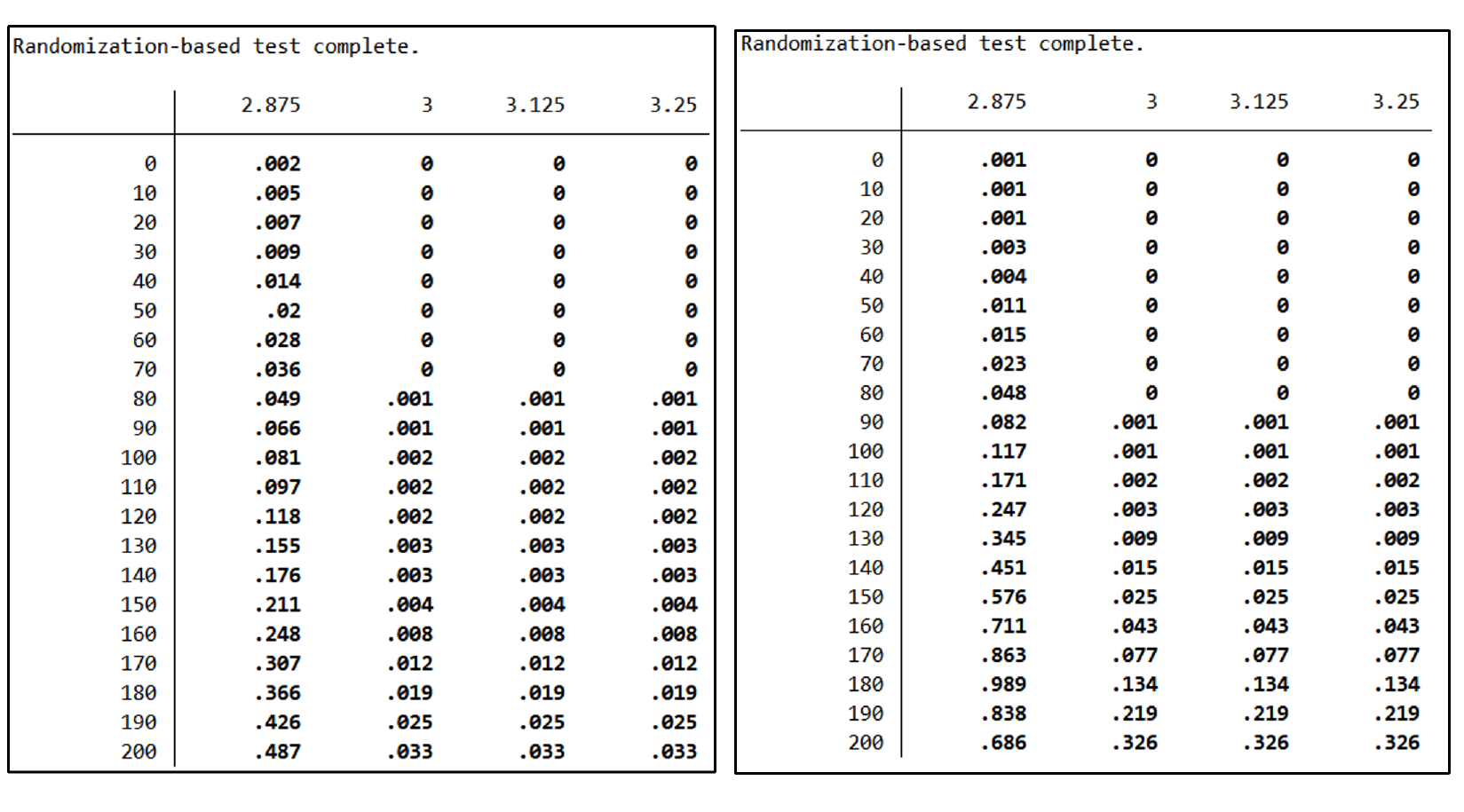}
		\label{winlr}
	}
	\vspace{2mm}
	\subfigure[Gini coefficient changes with window]{
		\includegraphics[width=0.8\linewidth]{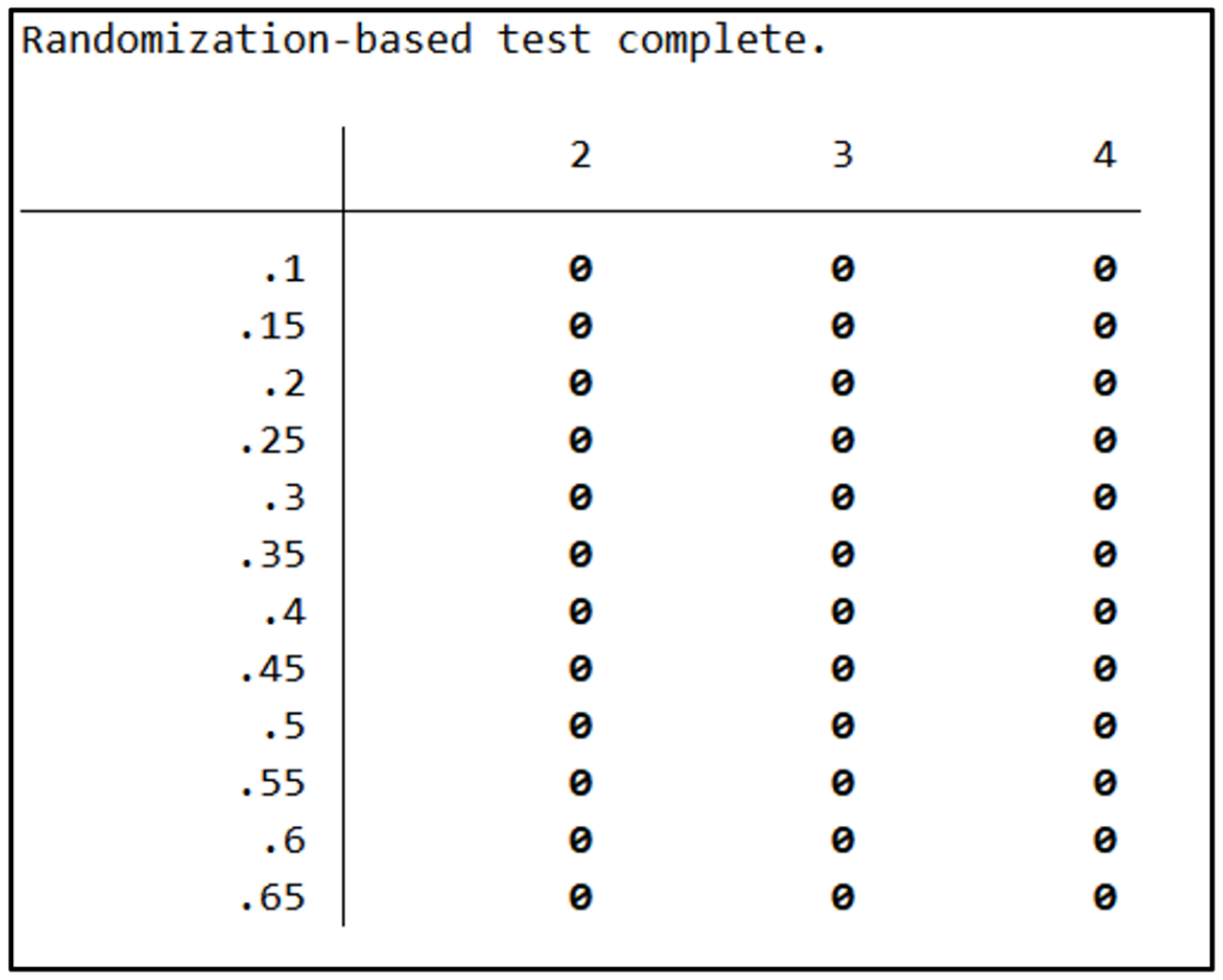}
		\label{winc}
	}
	\caption{\textbf{The three explained items change with the window}}
	\label{winx}
\end{figure}


%% file: figures/robust.tex
\begin{figure}[htpb]
	\centering
	\hspace{-1.5mm}
	\includegraphics[width=0.8\linewidth]{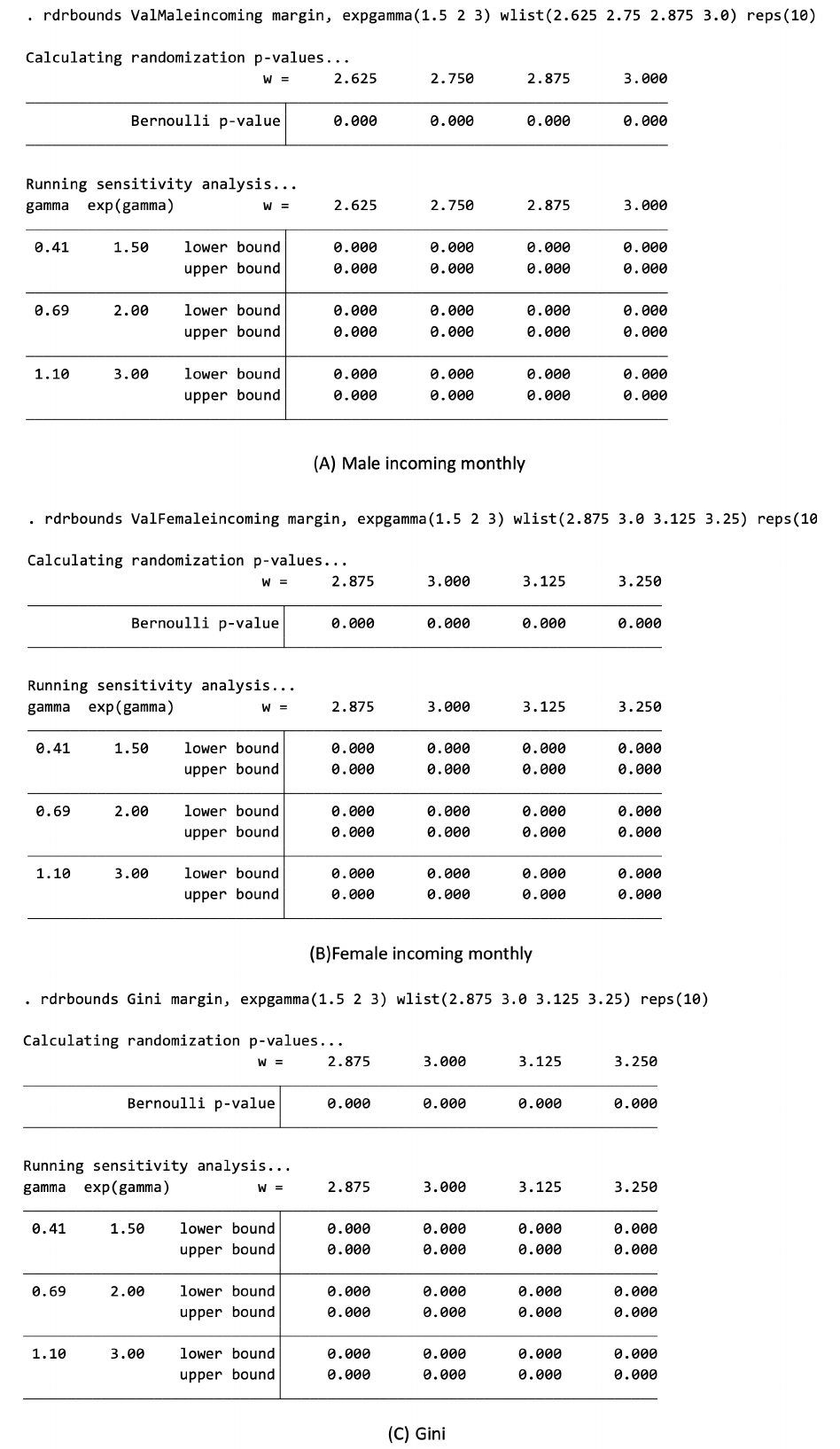}
	\caption{\textbf{Robustness test results for different test statistics}}
	\label{robust}
	\vspace{-.5mm}
\end{figure}
